**Title:**

A patient-specific approach for quantitative and automatic analysis of computed tomography images in lung disease: application to COVID-19 patients

**Authors:**

L. Berta[1], C. De Mattia[1], F. Rizzetto[2], S. Carrazza[5], P.E. Colombo[1], R. Fumagalli[3,4], T. Langer[3,4], D. Lizio[1], A. Vanzulli[2,6], A. Torresin[1] on behalf of the Niguarda COVID Workgroup.

[1] Department of MedicalPhysics, ASST Grande Ospedale Metropolitano Niguarda, Piazza Ospedale Maggiore 3, 20162 Milan, Italy;

[2] Department of Radiology, ASST Grande Ospedale Metropolitano Niguarda, Piazza Ospedale Maggiore 3, 20162 Milan, Italy;

[3] Department of Medicine and Surgery, University of Milan-Bicocca, Monza, Italy.

[4] Department of Anaesthesia and Intensive Care Medicine, ASST Grande Ospedale Metropolitano Niguarda, Piazza Ospedale Maggiore 3, 20162 Milan, Italy;

[5] Department of Physics, Università degli Studi di Milano and INFN Sezione di Milano, via Giovanni Celoria 16, 20133 Milan, Italy;

[6] Department of Oncology and Hemato-Oncology, Università degli Studi di Milano, via Festa del Perdono 7, 20122 Milan, Italy.

Corresponding author: Alberto Torresin: alberto.torresin@unimi.it

# Abstract

**Purpose**:

Quantitative metrics in lung computed tomography (CT) images have been widely used, often without a clear connection with physiology. This work proposes a patient-independent model for the estimation of well-aerated volume of lungs in CT images (WAVE).

**Methods**:

A Gaussian fit, with mean (Mu.f) and width (Sigma.f) values, was applied to the lower CT histogram data points of the lung to provide the estimation of the well-aerated lung volume (WAVE.f). Independence from CT reconstruction parameters and respiratory cycle was analysed using healthy lung CT images and 4DCT acquisitions. The Gaussian metrics and first order radiomic features calculated for a third cohort of COVID-19 patients were compared with those relative to healthy lungs. Each lung was further segmented in 24 subregions and a new biomarker derived from Gaussian fit parameter Mu.f was proposed to represent the local density changes.

**Results**:

WAVE.f resulted independent from the respiratory motion in 80% of the cases. Differences of 1%, 2% and up to 14% resulted comparing a moderate iterative strength and FBP algorithm, 1 and 3 mm of slice thickness and different reconstruction kernel.Healthy subjects were significantly different from COVID-19 patients for all the metrics calculated. Graphical representation of the local biomarker provides spatial and quantitative information in a single 2D picture.

**Conclusions**:

Unlike other metrics based on fixed histogram thresholds, this model is able to consider the inter- and intra-subject variability. In addition, it defines a local biomarker to quantify the severity of the disease, independently ofthe observer.

# Introduction

COVID-19 is a complex infectious disease characterized by common and non-specific symptoms, such as fever, cough, shortness of breath and fatigue, and a broad spectrum of clinical manifestation, ranging from asymptomatic infection to respiratory failure requiring oxygen support or invasive ventilation[1].

Computed tomography (CT) is the current standard of reference to assess lung alteration, even at early stage of the disease, when the patient has few or no symptoms, and to monitor the course of the disease at different time points[2,3].Despite the increase in chest CT examinations due to COVID-19 pandemic, the use of low dose protocols guarantees a very low risk of cancer induction[4].

The majority of COVID-19 studies use a *qualitative* approach, describing the lesions by visual and pictorial assessment. A lexicon for the description of chest CT imaging findings in coronavirus disease, *i.e.* the COVID-19 Reporting And Data System (COVID-RADS)[5], was proposed in addition to the terminology endorsed by the Fleischner Society Nomenclature Committee[6,7]. The purpose is to standardize terminology and communication and assessing a possible COVID-19 presence.

In addition, *visual quantitative analysis*, i.e. the scoring oflung abnormalities assessed by visual interpretation of CT images, and thedensitometric evaluations based on the histogram of the Hounsfield Unit (HU) distribution, were demonstrated useful to predict clinical severity[8–10].

Lung segmentation, the first step required for a quantitative analysis of medical images[11], can be performed nowadays through several fully automatic tools that help to reduce the intra and the inte-rreader variability[12,13].

Many indexes can be derived from the lung density histogram, starting from simple measurements, as the mean density value[14], to the measurement of the relative area of emphysema in patients with chronic obstructive pulmonary diseases[15]or the extraction of descriptive parameters of the histogram such as kurtosis and skewness[16].

Analysis of CT images has already been used in the past to better understand the pathophysiology of the acute respiratory distress syndrome [ARDS][17], using the density histogram to define lung compartments with different aeration.

In literature, different values of Hounsfield Units (HU)have been proposed to define these compartments[18]. To quantify the well-ventilated regions in COVID-19 patients, a potential surrogate to estimate the residual respiratory function and the alveolar recruitment during ventilation[19,20], the interval between -950 HU and -700 HU was proposed[21,22]. This threshold was already used in the past for studies involving other lung diseases[23,24], even if without a fully consensus, especially for ARDS. For example, a recent work used the density of -500 HU, already suggested by Gattinoni et al.[18], to discriminate between normally (-900, -501 HU) and compromised lung (-500, +100 HU), then further divided in poorly (-101, -500 HU)and non-aerated(-100, +100 HU) [25].

Despite being extremely informative, chest quantitative analysis has several drawbacks limiting its routine application.

First, the natural inter-patient and intra-patient variability, mainly due to the respiratory cycle. Several studiesperformed inspiratory and/or expiratory breath holds in combination with mechanical ventilation to standardize lung inflation during image acquisition[20,26]. In clinical practice, however, chest scans are obtained at the breathing-in point, inviting the patient to hold the breath. Holding the breath may however be a struggle for patients with a compromised status[27]and it is difficult to monitor respiratory phase without dedicated equipment.

Another limitation of current quantitative approaches is that the results are often not readily interpretable as physiological parameters but only as mathematical descriptors of the distribution of voxel valuesof a digital image.

Finally, acquisition and reconstruction CT parameters inevitably affect any kind of quantitative imaging analysis, with an impact on its usefulness which has to be evaluated for each individual source of variability[28].The use of a different reconstruction kernel than Visual CT (VCT) studies and the standardization of the protocols for the studies dedicated to Quantitative CT analysis (QCT) of the lungs was recommended by Neweel et al.[29].

The aim of this study was therefore to develop an automatic patient-customized lung analyzer to overcome the definition of aerated and pathologic regions based on thresholding of the density histogram, taking into account the patient variability and different CT acquisition protocols. Different reconstructions of the same scan series of healthy lung CT images, changing algorithm, kernel and the slice thickness, were used to test the reproducibility and the uncertainty of the proposed method. A dataset of 4DCT images, used routinely forguiding radiotherapy treatment planning[30],allowed to better understand the impact of the respiratory cycle on the density histogram and the related metrics. Finally, the analysis was applied on a COVID-19 cohort to distinguish the well-aerated and the pathological regions. A comparison with a method of CT lung histogram analysis based on pre-determined thresholds was added.

## Material and Methods

This retrospective study was approved by the Local Ethics Committee. The need for informed consent from individual patients was waived owing to the retrospective nature of the study.

**Study population**

We analyzed three cohorts of patients.
The first one included 20 patients (10 males, 10 females, median age 47.9 years, range 14-93 years), with a CT scan including the entire lung volume but not for pulmonary diseases (healthy lungs without alterations).

The exams were retrospectively collected in ASST Niguarda Hospital in July-August 2020 from Emergency Department CT scanner.

The second cohort was composed of 20 4DCT of locally-advanced non-small cell lung cancer patients (tumor volume: median 51cc, range 7-392 cc), publicly available in the Cancer Imaging Archive[31], acquired at the VCU Massey Cancer Center in the Department of Radiation Oncology, from 2008 through 2012, as a reference image for radiotherapy planning.

The third cohort was composed of 20 patients (17 males, 3 females, median age 58.5 years, range 33-73 years) randomly selected amongst those admitted in intensive care unit within the 48 hours after CT scan acquisition in March 2020 in Niguarda Hospital, with positive CT chest and positive Real-Time Polymerase Chain Reaction for SARS-CoV-2.

**CT protocol**

*Dataset 1*

CT studies of the first cohort were all acquired on a single CT scanner, a Somatom Edge unit, (Siemens AG, Forchheim, Germany) and with the same acquisition protocol. CT scans involving the entire lung were performed using a whole-body protocol with contrast agent and automatic exposure control (AEC) and automatic selection of the tube voltage (14 studies at 120 $kV_p$, 5 at 100 $kV_p$, 1 at 140 $kV_p$). We selected the basal CT phase, contrast not-enhanced, reconstructed for diagnostic intent (VCT series), using iterative algorithms (IR, Safire, S1), 3 mm as slice thickness and sharp kernel (Bl57).

*Dataset 1a*

Using the raw data of the first patient cohort, several sets of reconstructed images were obtained changing the slice thickness (1, 3, and 5 mm), the kernel (Bl57, Br38) and the strength of the iterative algorithm from pure Filtered Back Projection (FBP) to different level of SAFIRE blending (IR-S1, IR-S3, IR-S5).

*Dataset 2*

The second cohort 4DCT images were acquired on the CT 16-slice Brilliance Big Bore (Philips Medical Systems, Andover, MA), in helical mode at 120 $kV_p$, acquiring the respiratory signal trough the Real-time Position Management of the Varian Medical Systems. The raw data was sorting in 10 breathing phases, identified as a percentage, where the 0% phase corresponds to end of inhalation. Each 3D image was reconstructed with 3 mm slice thickness range using a soft kernel. All information regarding 4DCT protocol and patients of this database can be found in reference[32].

*Dataset 3*

For the third cohort, a CT protocol on the same scanner of dataset 1, with the same acquisition and reconstruction parameters, was used. Non-contrast chest CT scans were performed in supine position, during inspiratory breath-hold, within the limits of the collaboration status of the patient.

The scan parameters for each dataset are detailed in Table 1. Table 2 summarizes the CT reconstruction series added to the included exams.

| Dataset | Study | Number of patients (M/F) | Age (y) [Range] | Scanner Model | Vendor | Tube Voltage (kV) | ParametersStudied |
|---|---|---|---|---|---|---|---|
| 1 | Emergency access without pulmonary disease | 20 (10/10) | 48.5[14-93] | SomatomEdge | Siemens | 120 (14) 100 (5) 140 (1) | Fit parameters |
| 1a | Emergency access without pulmonary disease | 20(10/10) | 48.5[14-93] | SomatomEdge | Siemens | 120 | Kernel, slice thickness and algorithm |
| 2 | 4DCT | 20(ND) | ND | Brilliance Big More | Philips | 120 | Respiratoryphase |
| 3 | Emergency access with RT-PCR positive | 20 (17/3) | 58.5[33-73] | SomatomEdge | Siemens | 120 | Biomarkers for patients with compromised lung |

M: Male; F: Female; RT-PCR: Real-Time Polymerase Chain Reaction

Table 1: Demographic description of the datasets used and acquisition parameters of the CT series. For each dataset, the investigated variables were reported.

| Dataset | Study | Vendor | SliceThickness (mm) | ReconstructionAlgorithm | ConvolutionKernel |
|---|---|---|---|---|---|
| 1a | Emergency access without pulmonary disease | Siemens | 1 | Safire S1 | Bl57 |
| | | | 3 | Safire S1 | Bl57 |
| | | | 5 | Safire S1 | Bl57 |
| | | | 3 | FBP | Bl57 |
| | | | 3 | Safire S3 | Bl57 |
| | | | 3 | Safire S5 | Bl57 |
| | | | 3 | Safire S1 | Br38 |

Table 2: Details of the reconstructed series calculated from the raw data of the dataset 1.

**Images analysis**

*Lung segmentation*

Anonymized datasets were exported to a dedicated workstation where, through the extension module Chest Imaging Platform (CIP, Applied Chest Imaging Laboratory; Boston, Massachusetts, USA) of the open-source 3D Slicer software (version 4.10.2, https://www.slicer.org)[33], a fully automatic segmentation of the lung was performed. For each patient we segmented the first reconstruction series and then we applied the same mask at different slice thickness and kernel. For the dataset 2, each 3D series representing a phase of the 4D acquisition was segmented.

Lung segmentations of COVID-19 dataset images were carefully reviewed by an experienced radiologist and manually corrected where automatic algorithm failed.

CIP module automatically distinguishes the right and the left lung, and provides a subdivision of lung in upper, middle and inferior regions, using anatomical landmarks. As a first step, lung analysis was performed considering the whole pulmonary volume.

Subsequently, the imaging features of the COVID-19 dataset were analyzed and interpreted in detail through the further segmentation of lung regions in subregions. We added the ventral-dorsal and medial-distal subdivision, for a total of 24 subregions.

In each axial slice of the upper and middle lung region, the line connecting the centroids of the right and of the left lung was used to separate ventral and dorsal regions. The distal and medial regions were identified by the perpendicular lines passing through the centroids of each lung [Fig.1]. For the inferior region, the subdivision was calculated extending the results of the lowest slice in the middle lung region.

*Well-Aerated Volume Estimation*

A dedicated software was written in JavaScript language to automatically calculate and analyze the histograms of the CT images within ImageJ environment[34]. This software needs as input a 3DCT chest acquisition and applies the mask of the lung and its subvolumes. All the following analysis concerns only the voxels identified by the lung mask. A relative frequency histogram of HU values is then calculated in the range $-1020 \div +300$ HU with bin width chosen by the user.

The proposed method estimates the well-aerated volume (WAVE) under the assumption that the distribution of the HU values of the voxels regarding exclusively *healthy* parenchymal tissue would be described by a Gaussian function[35]. In fact, the healthy parenchymal tissue can be seen as a mix of air and water arranged in cellular architecture with an average uniform density. Furthermore, due to the random nature of cellular architecture, when measured at a voxel scale, the density is no longer constant. Indeed, healthy lung densitometry has a distribution of different values and the resulting density histogram would be similar to a Gaussian distribution function. If the main source of noise in CT images is due to quantum noise, the expected distribution of HU values in cellular material is still described by a Gaussian curve.

The lung segmentation encompasses mainly the healthy parenchyma, but other structures are included as well. For instance, vessels and/or lesions are frequently included in the region of interest. These structures, due to their higher physical density, extend the tail of the histogram towards higher values of HU. However, if the altered tissue covers a limited volume, the left side of the histogram could still be representative of healthy lung tissue only and have a Gaussian shape centered on the histogram modal value.

Based on these assumptions, WAVE.f is defined by integrating between -1020 to +300 HU the Gaussian function with formula:

$$y(x) = Height.f * e^{\frac{-(x-Mu.f)^2}{2*Sigma.f^2}} \quad (1)$$

used to fit the points around the modal value in a CT lung histogram.

Mu.f and Sigma.f are the central and the standard deviation values of the Gaussian curve, respectively, and Height.f is a normalization factor, not relevant in this work.

Two approaches were used to obtain the fit parametersusing the Curve Fittter Toolimplemented in ImageJv.1.53asetting 6000 as maximum iterations number, 2 as number of restarts and $10^{-10}$ as error tolerance. A polynomial fit of the second order was used to fit the logarithm of the HU histogram frequencies with the formula:

$$\ln(y) = a + bx + cx^2 \quad (2)$$

Alternatively, the fit parameters were calculated with the using the "Gaussian (no offset) function", readily available between the fitting functions.

The relationships between parameters of equations(1) and (2) are the following:

$$Heigt.f = e^{(a-\frac{b^2}{4c})}$$

$$Mu.f = -\frac{b}{c}$$

$$Sigma.f = \sqrt{\left|-\frac{1}{2c}\right|}$$

.

The fitting range was defined as the HU points in the histogram included within a starting point, on the left of the modal value, and an end point, on the right of the modal value, identified by their relative frequency values (Fig. 2).

We set: 3, 5, and 10 as bin width in the histogram; 20%, 30% and 40% as percentage of the modal value for the starting points and 80%, 85% and 90% of the modal value for the end point. A total of 27 combinations of bin width and range values were explored for the Gaussian fitting of the CT lung histogram points on the dataset 1 in order to evaluate the WAVE.f dependence on these parameters.

The dataset 1a was used to evaluate the dependency of the Mu.f, Sigma.f and WAVE.f values as a function of slice thickness, kernel and reconstruction algorithm.

For comparison, this variability was assessed on another lung metric, previously described by several authors[11,15,18,21,25], calculated integrating histogram within fixed threshold in the HU range -950÷-700, hereafter called WAVE.th. Dataset 2 allowed to investigate the correlation between the Gaussian curve metrics and the respiratory phase. Different level of inspiration can affect the central value (Mu.f) of the Gaussian function but the area under this curve should represent an estimation of the well-aerated lung volume, independently of the respiratory phase.

The tidal volume (TD) was calculated as the maximum difference in lung volume during the respiratory cycle and patients with TD lower than 390 mL were excluded from dataset 2[36,37]. With respect to tumor location, contralateral lungs of the remaining patients, were analyzed and the metrics obtained at the various respiratory phases were correlated with the percentage variation of the lung volume.

*Biomarkers in COVID-19 patients imaging*

For dataset 3, in addition to all the metrics described above (Mu.f, Sigma.f, WAVE.f, WAVE.th), we calculated also standard first-order radiomic features (i.e. average HU, Skewness, Kurtosis and 0.25, 0.5, 0.75 and 0.90 percentiles) from the CT histogram of the whole lung volume. For comparison of healthy and affected lungs, the same biomarkers were calculated also in patients of dataset 1.

A new biomarker was introduced to assess local change in lung density with respect to the healthy parenchyma and defined as:

$$\Delta HU_{Mu-Avg_i} = \overline{HU}_i - Mu.f \quad (3)$$

In equation (3), Mu.f is the fit parameter relative of the histogram for the entire lung whereas $\overline{HU}_i$ is the average value of voxels in a specific region i of the lung. This biomarker was then calculated in each of the 24 lung subregions of COVID-19 patients.

For each patient, the range and the median values over all subregions were considered as a metric related to the severity of the disease. We calculated the average and the standard deviation values over the 20 patients.

**Data analysis**

Statistical analysis was performed by using the Real Statistics Resource Pack software (release 6.8, www.realstatistics.com). Saphiro-Wilk and Levene tests were used to assess the normality of the distributions and the homogeneity of the variance.

*Dataset 1*

Theparametersof the Gaussian function were calculated withboth the polynomial andthe exponential equations, fitting data points of histogram calculated with 5 HU using a 30% and 90% of the modal value as starting and ending points. The two modelswere compared using the relationship of equation (3).Differences between Height, Mu.f, Sigma.f and WAVE.f values resulting from the polynomial and exponential models wereassessed for each healthy subject of dataset 1 and their correlation was assessed using the Pearson coefficient.

To estimate the statistical significance of the differences in WAVE.f calculation due to the bin width and the range for the Gaussian fitting, analysis of variance (ANOVA) was performed (significance level of 0.05), changing one parameter at a time and keeping the other two fixed.

*Dataset 1a*

ANOVA was applied (significance level of 0.05) to the Mu.f, Sigma.f, WAVE.f and WAVE.th. values calculated in CT images reconstructed from the same sinogram changing the reconstruction parameters (slice thickness, reconstruction algorithm, reconstruction kernel).
Paired t-tests were then calculated between metrics calculated from the VCT (3 mm as slice thickness, IR-S1 and Bl57 as reconstruction algorithm and kernel) and the other post-processed series.
A p value less than 0.05 was considered indicating a significant difference.

*Dataset 2*

Pearson correlation coefficient was calculated to estimate the linear correlation between the metrics described (Mu.f, Sigma.f, WAVE.f, WAVE.th) and the lung volume, expressed as a percentage of the maximum volume, at the 10 different phases of the respiratory cycle. According to the distribution of the Pearson correlation coefficient with n=10, values above 0.632 and 0.765were considered indicating significant and strongly significant correlation, respectively (significance level: 0.05 and 0.01). McNemar test was used to assess the significance of the differences in the number of patients with WAVE.f and WAVE.th correlated with the respiratory phase.

*Dataset 3*

A t-test was used to compare results of this dataset with those obtained in dataset 1 (significance level of 0.05). A description of lung disease is given by a graphical representation of the CT lung histogram metrics in the 24 lung subregions, hereafter called lungogram.

# Results

The properties and results of the metrics and biomarkers (WAVE.f, WAVE.th and $\Delta HU_{Mu-Avg}$) were studied in relation to the specific datasets described above.

*Dataset 1*

Average differences forHeight.f,Mu.f, Sigma.f and WAVE.f calculated with polynomial and exponential modelwere0.1%, 0.2 HU and 0.3 HU and 0.3%, respectively.Pearson coefficients for these metrics were 1.00, 1.00, 0.979 and 0.988. Since CT histogram are generally reported in linear scale, hereafter, only results of the exponential fitting will be reported.

Fitting parameters, such as bin width, starting point and end point of the fitting range, did not affected significantly the WAVE.f calculation. According to the ANOVA tests, the p-value was greater than 0.812 for all them.

In our analysis a value of 5 HU as bin width was selected since it provided a trade-off between smoothness and resolution in the histogram.In healthy subjects, the choice of starting and ending point is not relevant, since CT histogram of the healthy parenchyma does not deviate importantly from a Gaussian distribution.In this work, considering the application on compromised lungs, 30% on the left side of the modal histogram value was set as starting point (see the point (a) of fig. 2). In this way, the fitting range should avoid the less dense part of the histogram possibly due to the presence of emphysema (Fig.2). As ending point, 90% on the right side of the modal histogram value was set (see the point (b) of fig. 2). This choice should reduce the use of the points on the right side of the histogram in patients with increased lung density, as occurs in particular for the COVID-19 dataset.

*Dataset 1a*

ANOVA found statistically significant differences changing slice thickness and iterative strength for the estimation ofWAVE.f ($p<0.001$, $p<0.001$), WAVE.th ($p<0.001$, $p<0.001$) and Sigma.f ($p=0.026$, $p<0.001$). For Mu.f the differences were not statistically significant ($p=0.629$, $p=0.896$). All results regarding analysis of dataset 1a are summarized in Table 3.

The Gaussian fit parameter Mu.f did not showed significant differences in paired t-test between the standard reconstruction (VCT series) and the other series comparing 3 and 5 mm as slice thickness or IR-S1 and IR-S3 as reconstruction algorithm (Table 4). Sigma.f and WAVE.th had always significant differences changing the reconstruction parameters, while WAVE.f did not differ using slice thickness lower than 3 mm.

| Kernel and reconstructionalgorithm | SliceThickness (mm) | Mu.f (HU) | Sigma.f (HU) | WAVE.f (%) | WAVE.th (%) |
|---|---|---|---|---|---|
| Bl57 IR-1 | 1mm | -887 (±26) | 99 (±10) | 82 (±2) | 66 (±3) |
| Bl57 IR-1 | 3mm | -868 (±26) | 74 (±11) | 84 (±2) | 75 (±4) |
| Bl57 IR-1 | 5mm | -867 (±25) | 62 (±7) | 80 (±2) | 77 (±3) |
| | | | | | |
| Bl57 FBP | 3mm | -870 (±26) | 83 (±11) | 83 (±2) | 72 (±3) |
| Bl57 IR-1 | 3mm | -868 (±26) | 74 (±11) | 84 (±2) | 75 (±3) |
| Bl57 IR-3 | 3mm | -866 (±25) | 51 (±7) | 80 (±3) | 82 (±3) |
| Bl57 IR-5 | 3mm | -864 (±24) | 34 (±6) | 77 (±4) | 87 (±2) |
| | | | | | |
| Bl57 IR-1 | 3mm | -868 (±26) | 74 (±11) | 84 (±2) | 75 (±4) |
| Br38 IR-1 | 3mm | -852 (±25) | 30 (±7) | 72 (±6) | 88 (±3) |

Table 3: Mean and Standard Deviation of Mu.f, Sigma.f, WAVE.f and WAVE.th values calculated for the dataset 1a, changing the reconstruction parameters (slice thickness, reconstruction algorithm, kernel).

| | | Mu.f | Sigma.f | WAVE.f | WAVE.th |
|---|---|---|---|---|---|
| SliceThickness | 1 vs 3 mm | <0.001 | <0.001 | 0.09 | <0.001 |
| | 1 vs 5 mm | <0.001 | <0.001 | <0.001 | <0.001 |
| | 3 vs 5 mm | 0.8 | <0.001 | <0.001 | <0.001 |
| Reconstruction Algorithm | FBP vs IR-S1 | 0.04 | <0.001 | 0.01 | <0.001 |
| | FBP vs IR-S3 | <0.001 | <0.001 | <0.001 | <0.001 |
| | IR-S1 vs IR-S3 | 0.2 | <0.001 | <0.001 | <0.001 |
| | IR-S1 vs IR-S5 | 0.03 | <0.001 | <0.001 | <0.001 |
| Kernel | Bl57 vs Br38 | <0.001 | <0.001 | <0.001 | <0.001 |

Table 4: P-values results of paired t-test on dataset 1a changing the reconstruction parameters (slice thickness, reconstruction algorithm, kernel), keeping as reference the standard series (3 mm, IR-S1, Bl57).

*Dataset 2*

From the second cohort, 5 patients were excluded for their limited difference in lung volume in the respiratory cycle (TD <390 mL). For the remaining 15 patients, the results of the correlation between the calculated metrics and lung expansion are reported in Table 5. WAVE.f resulted correlated significantly in 3 and strongly significant in 1 out of 15 cases. As expected, significant correlation was found for Mu.f in all cases. Sigma.f resulted significantly correlated in 11 cases . For WAVE.th, the correlation was strongly significant for 12 patients .

McNemar test applied to the number of cases reported in Table 5 for WAVE.f and WAVE.th returned p-values of 0.009 and 0.001 for significant and strongly significant correlation with respiratory cycle indicating a significant difference between the two metrics in the correlation with the respiratory cycle.

The effect of respiratory cycle on the lung metrics calculated in CT images for two patients (P104, P111) from the dataset 2 representative of results of Table 5 is reported in Fig.3. Linear regression lines are displayed when a strongly significant correlation was found between lung volume and lung biomarkers (Mu.f, Sigma.f and WAVE.f and WAVE.th).

| Alpha | Pearson coefficient | | Mu.f | Sigma.f | WAVE.f | WAVE.th |
|---|---|---|---|---|---|---|
| 0.05 (significant correlation) | > 0.632 | n | 15 | 11 | 3 | 12 |
| 0.01 (strongly significant correlation) | > 0.765 | n | 14 | 8 | 1 | 12 |

Table 5: Number of cases with lung metrics calculated in 4DCT images, correlated with respiratory phase.

*Dataset 3*

Metrics calculated for the whole lung in COVID-19 patients of dataset 3 were compared with results of normal lung patients of dataset 1. Boxplot of the lung metrics calculated in dataset 1 (healthy lung) and dataset 3 (COVID-19 patients) are reported in Fig.4. Differences were significant for all the metrics. A p-value of 0.007 was found for Mu.f comparison while for all the other metrics p-values were <0.001. In healthy subjects average (standard deviation) values for WAVE.f and WAVE.th resulted 84% (2%) and 75% (4%). Results of the biomarker $\Delta HU_{Mu-Avg}$ calculated in the 24 subregions for the dataset 3 are summarized in Table 6. The average (standard deviation) values for range and median of $\Delta HU_{Mu-Avg}$ were, respectively, 411 (127) HU and 179 (59) HU. Examples of analysis performed for two patients, including histogram, Gaussian fit and lungogram, are reported in Fig.5. The difference between CT lung histogram and Gaussian fit data, indicated with the blue line, represents the distribution of all non-healthy lung parenchyma like vessels, airways and disease-involved tissue.

|  | $\Delta HU_{Mu-Avg}$(HU) | | | |
|---|---|---|---|---|
| **Patient ID** | **Minimum** | **Maximum** | **Range** | **Median** |
| P_01 | -3 | 231 | 234 | 76 |
| P_02 | 87 | 301 | 214 | 165 |
| P_03 | 55 | 528 | 472 | 299 |
| P_04 | 31 | 422 | 390 | 202 |
| P_05 | 26 | 293 | 267 | 120 |
| P_06 | 28 | 331 | 304 | 155 |
| P_07 | 41 | 666 | 625 | 260 |
| P_08 | 18 | 352 | 334 | 115 |
| P_09 | 73 | 372 | 299 | 144 |
| P_10 | -11 | 467 | 479 | 155 |
| P_11 | 60 | 617 | 557 | 215 |
| P_12 | 26 | 339 | 313 | 144 |
| P_13 | 14 | 647 | 633 | 117 |
| P_14 | 14 | 472 | 458 | 177 |
| P_15 | 49 | 386 | 337 | 216 |
| P_16 | 42 | 480 | 438 | 175 |
| P_17 | 30 | 522 | 493 | 197 |
| P_18 | 54 | 644 | 590 | 284 |
| P_19 | 49 | 396 | 347 | 232 |
| P_20 | 14 | 446 | 432 | 144 |
|  |  |  |  |  |
| Average | 35 | 446 | 411 | 179 |
| Standard Deviation | 25 | 127 | 127 | 59 |

Table 6: Minimum, maximum, median values and range of the biomarker $\Delta HU_{Mu\text{-}Avg}$ calculated over the 24 subregions for all patients of the dataset 3. The last two rows report the average and the standard deviation over the all dataset.

# Discussion

In this work a patient-specific and automatic model, based on a Gaussian fit of relative CT lung histogram, was described to quantify aerated and pathologic lung regions in healthy controls and in patients with interstitial pneumonia caused by SARS-COV-2.The parameters Mu.f and Sigma.f, which characterize the Gaussian model, are related to lung inflation and noise in the images of the healthy parenchyma, resepctively. WAVE.f is a new metric suitable for quantification of the well-aerated lungs that showed lower variability with physiological and technical factors than others alreadyproposed.

CT lung histogram fitting has already been reported in literature and metrics derived from histogram analysis have been used in the past. Obert et Al.[38]proposed a model in which cumulative CT lung histogram data were fitted with a logistic growth function in order to classify normal and pathological lungs. Despite the good results, this logistic model had a principle mathematical approach. In contrast, the model we propose takes into account the physical properties of the pulmonary parenchyma and the nature of noise of CT images.In this work, we used a non-linear equation to calculate the Mu.f and Sigma.fsince Gaussian fit is readily available withinthe program used to develop our model. Alternatively, a second order polynomial fitting of the logarithm of the histogram frequencies could be used making the procedure more independent from the platform used. However, the difference between the WAVE.f results was negligible for clinical purposes and the exponential fitting offers the advantage of having a clear graphical interpretation of CT lung histogram.

Usually, quantitative lung analysis applies fixed HU thresholds in order to divide lung volume in compartments with different aerations. However, a fixed HU range does not take in account the inter-patient variability or the phase of the respiratory cycle. This fact is highlighted in the examples reported in Fig. 5, where the Gaussian peak that accurately fits data points around the histogram modal value is close to the opposite limits of the fixed range (Mu.f=-765 HU for Case C, Mu.f=-911 for Case D). The inter-patient variability of the histogram modal value, which results in this asymmetry with respect to the fixed HU range, can be ascribed to clinical or physiological factors which only a patient-specific metric can take into account.

The model of this study was tested in two cohorts of healthy and critically ill patients, representing two opposite conditions of lung status. Under the hypothesis formulated here, a Gaussian curve describes the radiological aspects of healthy lung tissue: the central value represents the average density of the well-aerated tissue and the width includes all sources of image noise due to reconstruction parameters and anatomical texture. The two model parameters Mu.f and Sigma.f inevitably depend on the histogram data points used for fitting but the resulting WAVE.f values are independent of the fitting range used, at least in healthy lungs. Even in cases with severe lung impairment due tointerstitial pneumonia, the modal value of the histograms was representative of healthy lung tissue and the Gaussian fit properly represents the estimation of the ventilated lung as it takes into account the intra-patient variability. However, in a limited fraction of subjects havingmost of the lung volume compromised by opacities or solidifications,the histogram modal value of the whole lung can results with unusual values, not included in the range tipically characterizing the health lung tissue, i.e. between -950 and -700 HU . In these cases, the Gaussian model may not be directly applicable as the fit on the selected histogram data could return for Mu.fvalues that are not representative of the healthy tissue.

To take this into account, the definition of the fit range should not consider the modal value of the whole lung histogram but the presence of relative maxima in a limited HU range.This is evident in the example shown in figure 6, where the modal value of the histogram is greater than -700HU. The value of Mu.f obtained fitting data points around the modal value, or even in a limited range as shown in figure 6-A, is higher than -700 underlining the non-applicability of the model under these conditions.

On the other hand, in the case of patients with emphysema[39], because of the phenomenonknown as "air trapping", the Mu.f values might be underestimated if the fit range values are not adequate.To overcome the limitations of these specific cases, it is possible to select the histogram points to be fitted in a more tailored way, for example using additional conditions on the first or second derivative (Fig. 6-B). Alternatively,useful information could come from histograms of sub-regions where the diseaseis less prevalent and the modal value is representative of the healthy tissue, such as in the sub-regions of the same patient in Figure 6-C where the Gaussian model can be applied.

Regardingthe robustness, the metrics for QCT assessment should be independent from technical and physiological bias: analysis of dataset 1a and dataset 2 was aimed at studying the relationship of the metrics with respiratory cycle and with image reconstruction parameters.

The sources of variability in QCT for lung analysis are well known and reported in literature[40–42]. A protocol standardization is recommended to reduce results variability, but it is achievable only in prospective studies. Neither WAVE.f nor WAVE.th metrics resulted completely independent from all the reconstruction parameters. However, no significant differences were found in WAVE.f values calculated in images with slice thickness in the range 1-3 mm, the most used values for high resolution lung CT protocols.

An increase of slice thickness, as well as an increase of the strength of iterative algorithm, implies a decrease in image noise magnitude that impact systematically on the CT lung histograms and on the derived metrics. Comparing WAVE.fand WAVE.th values calculated in images reconstructed with different parameters, an overall lower variability was observed in the Gaussian model's metric due to its intrinsic ability to fit the actual data in the histogram.

A significant difference was found in all comparison when different algorithms were selected. The nonlinear effects of SAFIRE algorithm may explain the increase of differences found with the increase of the IR strength[43]. Nevertheless, the most limited difference in the WAVE.f results was found comparing a moderate strength of iteration and FBP algorithm. For this reason, VCT series, using Bl57 IR-S1 and 3mm as slice thickness, was considered suitable also for quantitative purpose.

Replacing the Bl57 kernel with the medium smooth Br38 kernel, not generally used for diagnostic purposes in chest imaging, implies the greatest differences in WAVE.f. However, we added this reconstruction series since images in public dataset 2 were available only with a soft kernel but medium-smooth kernel are not generally used for VCT of the lung parenchyma.

Another challenging task in medical imaging is organ motion[44–46]. In particular, for quantitative imaging assessment, the differences in lung volume due to the respiratory cycle result in different values of tissue density. This is clearly visible in Fig.3 and in Table 5, where the correlation for Mu.f was always significant. Sigma.f also showed a significant correlation in most of the cases, affecting the values of WAVE.th that, as consequence, showed a strongly significant correlation with lung inflation in 12 out of 15 of analyzed cases.On the other hand, WAVE.f resulted more stable with respiratory phases than other metrics. This is due to the customized properties of the proposed model that takes into account the inter- and intra- subject variability. It must be stressed that these results refer to radiotherapy patients, specifically trained to follow a shallow and regular breathing, while during diagnostic chest examination, the CT scan is performed by asking the patient to have deep breath, compatibly with his pathological state.

The estimation of the fractional volume of well aerated lung is calculated for both WAVE metrics from the integration of the CT lung histogram but with a substantial methodological difference: while the WAVE.th metric uses fixed HU range to identify the histogram area corresponding to the healthy lung parenchyma, in the proposed method the integration range is chosen according to an image-specific model. These different approaches are represented in figure 6-Band6-C. The light red shadowed areas under the curves represent the lung volume classified as "well aerated" according to the Gaussian model but not according to theWAVE.th definition. Analogously, the light blue shadowed areas under the curves represent the lung volume classified as "well aerated" according to the fixed threshold model but not according to the Gaussian metric. When the differences between these volumes are not compensated, a difference between the two WAVE metrics occurs. The relationship between WAVE.f and WAVE.th is shown in figure 7. As expected, lung opacifications and solidifications in COVID-19 patients reduce WAVE values for both metrics, with WAVE.th systematically lower than WAVE.f.Moreover, in healthy subjects no correlation exists becausethe variability of WAVE.th, that do not consider the actual respiratory phase, is not hidden by the inter-patient disease variability.

In healthy subjects, an average value of 84% (range: 79%-86%) was found for WAVE.f. This result represents the percentage of healthy parenchymal tissue of the entire lung volume in non-pathological lung images. Although the assessment of the accuracy of WAVE metrics was not amongst the aims of this work, a correspondence was found with results of morphometric studies. Townsley reported an average value of 84% (range: 77-87%) as fraction of overall anatomic lung defined as parenchyma[47]. By contrast, WAVE.th calculated in the same cohort of healthy subjects, showed lower results and a higher variability.

In the comparison between results of datasets 1 and 3, for all the metrics the t-test showed significant differences.The discrepancy between the two cohorts found for Mu.f can be explained as the limited capabilities of deep breath holding in patients with severe lung impairments that result in higher density values of parenchymal tissue.

The other radiomic features calculated from histograms (HU mean, Skewness and Kurtosis) have similar trends as reported in literature to distinguish healthy and pathological lungs[16,48].

As expected, also WAVE metrics clearly discerned the two cohorts, but its values, unlike the first-order radiomic features, can give quantitative information about the well-aerated lung that can be useful in the management of patients with ARDS[20].

Moreover, WAVE.f values are referred to the entire lung but interstitial pneumonia from SARS-CoV-2 produces solidifications and opacifications distributed heterogeneously in the lung parenchyma with different level of severity. Therefore, local changes in lung density can be assessed calculating $\Delta HU_{Mu\text{-}Avg}$ in any specific lung region. A graphical representation of this biomarker in the lungogram provides spatial and quantitative information of the patient's lung status in a single 2D picture of easy interpretation and suitable for clinical decision or inter patients' comparison.

This work has some limitations. First of all, the model is not suitable for all cases. When the disease affects most of the lung volume and the modal value of the CT numbers is shifted towards unexpected high values, a different approachmust be used to extend the applicability of the model to the entire lung. Possible solutions includedifferent criteria for the definition of the fit range or taking into account the histogram of the lung sub-regionsFurthermore, the evaluation of the impact of the reconstruction parameters on lung metrics was performed using only one scanner. However, such a study regarding both reconstruction and acquisition parameters is feasible only with physical phantom or with a very high number of cases involved.

# Conclusion

A method to analyse CT lung images based on the Gaussian fit of the histogram data has been developed and characterized.

In healthy lungs, WAVE.f, a new quantitative metric derived from physics assumptions and with physiological significance, demonstrates lower dependencies from technical or physiological parameters with respect tothe already reported equivalent metrics and its values were in good agreement with morphometric studies.

The complex patterns of lung diseases, such as those resulting from SARS-CoV-2 pneumonia, can be described by appropriate metrics calculated locally. The biomarker $\Delta HU_{Mu\text{-}Avg}$ is an absolute measure in Hounsfield Units of lung density and its values calculated in 24 lung subregions of COVID-19 patients combines quantitative and spatial information.

Finally, a validation of WAVE metrics is mandatory before its use for clinical decision. A future work using a larger sample of clinical images and functional data can be addressed to verify the hypothesis on which this model is built and to assess accuracy of the WAVE.f.

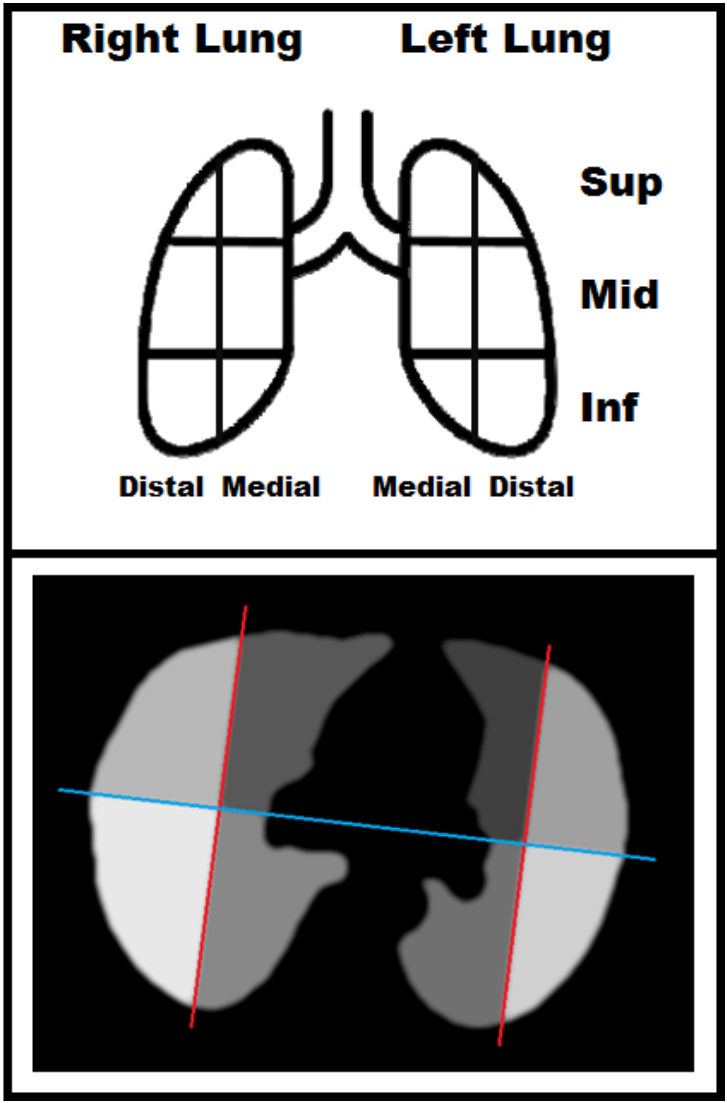

Fig.1: Graphical representation of the lung volume subdivided into: left, right, superior, middle, inferior, ventral, dorsal, medial, distal, for a total of 24 sub-volumes.

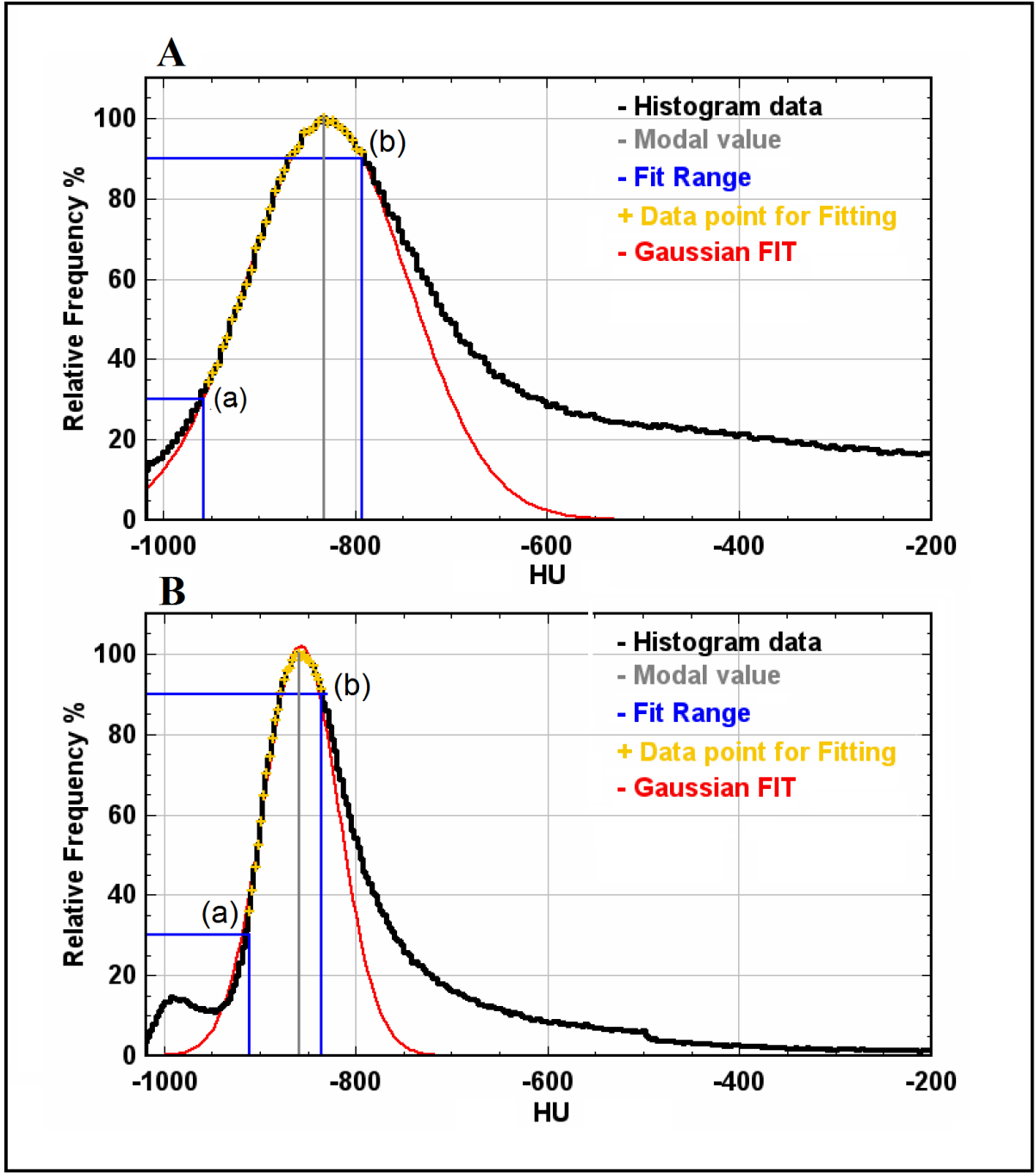

Fig.2: Examples of histogram analysis. Data are normalized to the modal value which is indicated with a vertical gray line. The black line represents the total lung histogram; the blue lines show the extremes of the range of the data points used for fitting, reported as orange crosses. Example A), CT lung histogram of a COVID-19 patient. Example B) CT lung histogram of a patient with emphysema from dataset 2.

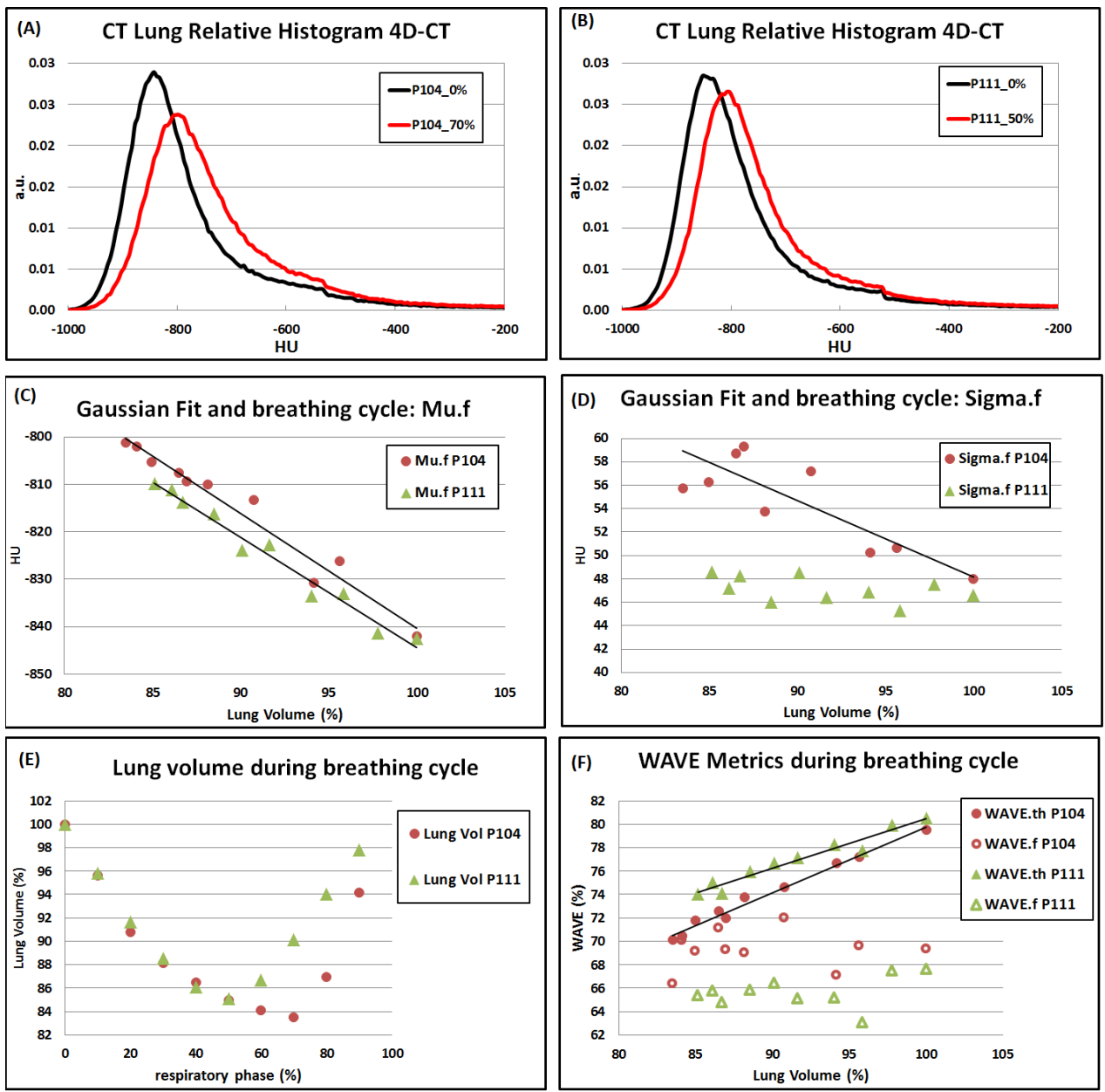

Fig.3: The effect of respiratory cycle on the lung metrics calculated in CT images for two patients from the dataset 2 representative of results reported in Table 5.

(A) and (B): CT lung relative histogram of the two patients in minimum and maximum inhale phases.

(C) and (D): fit parameters Mu.f and Sigma.f are plotted versus the lung volume

(E) : lung volume plotted versus the respiratory phase.

(F): WAVE metrics plotted versus lung volume

Linear regression lines are displayed when strong correlation was found

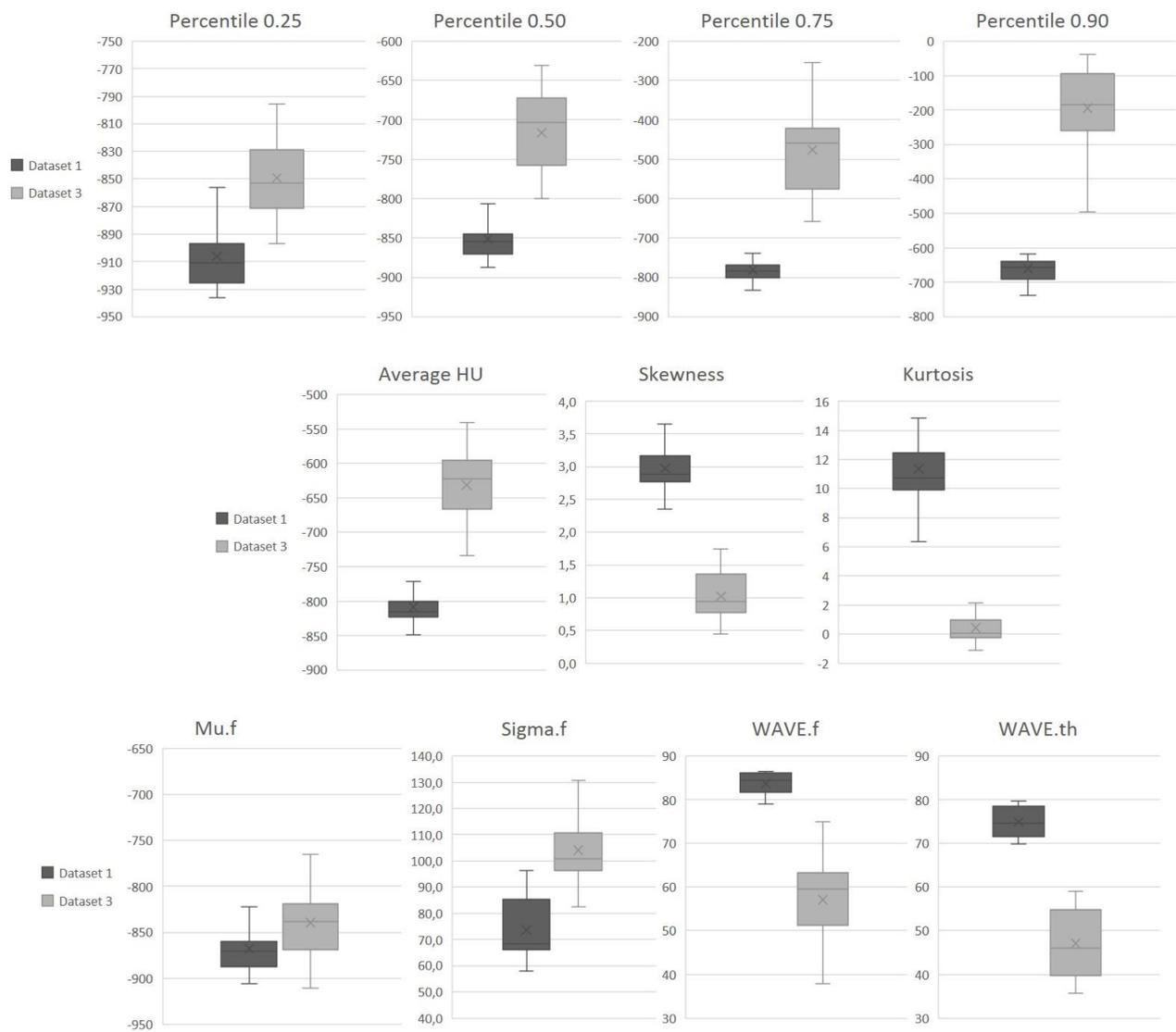

Fig.4: Boxplot of the lung metrics calculated in dataset 1 (normal lung) and dataset 3 (COVID-19 patients).

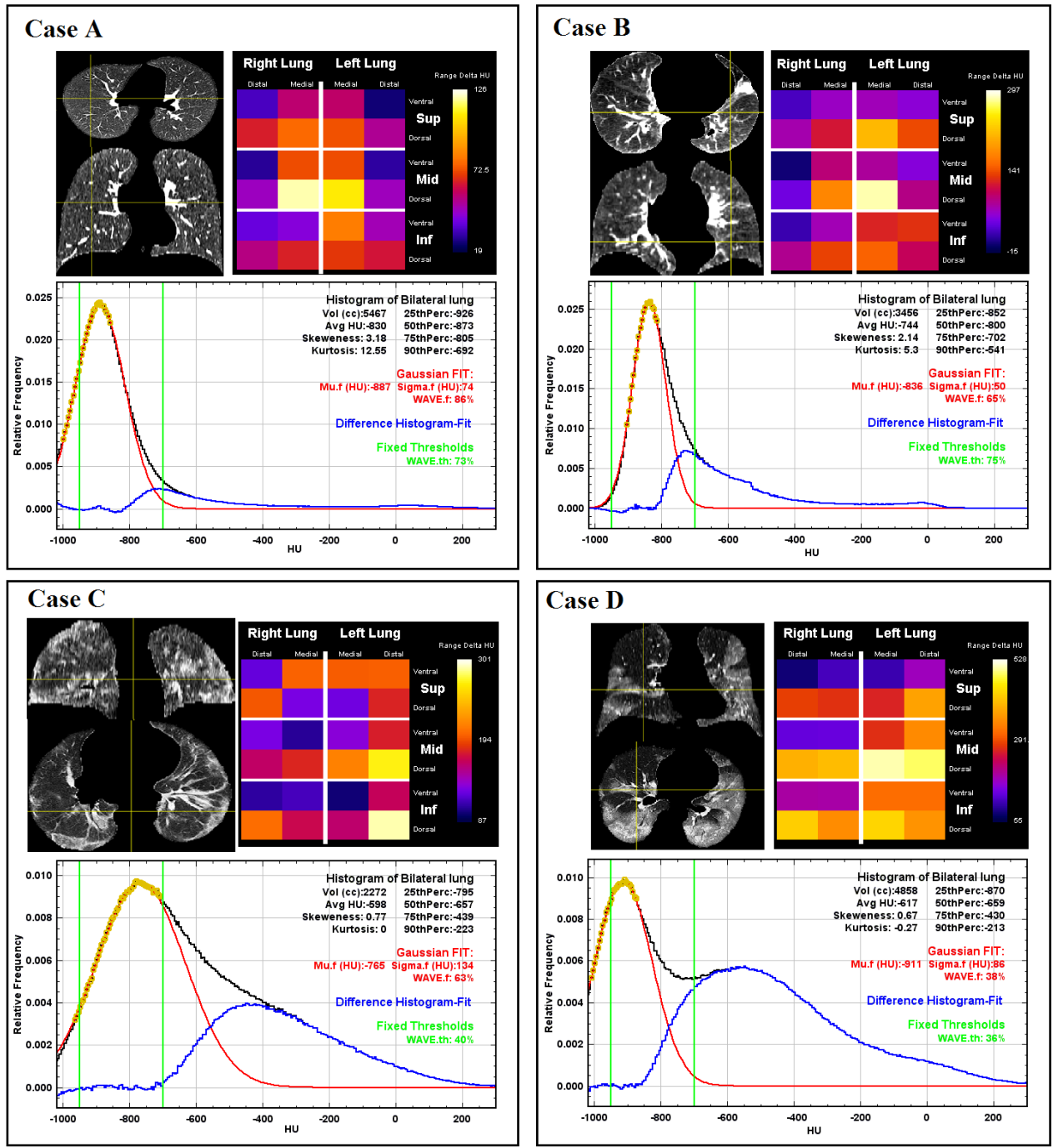

Fig.5: Examples of patients analyzed with the methods described in the materials and methods section. Segmented lungs are displayed in axial and coronal views. Local changes in density are reported in lungograms. In the plots, black curve represents the CT relative histogram of the entire lung. Yellow dots are the data points selected to calculate Gaussian fit (red curve). The difference between CT lung relative histogram and Gaussian fit is reported in blue while the HU range used for WAVE.th calculations is reported by two green vertical lines. The relative metrics are reported with the same colors as the curves in the graphs on which they are calculated. Lung histogram features (HU average, skewness, kurtosis, and percentiles) are reported in black,

Gaussian fit parameters (Mu.f, Sigma.f) and WAVE.f are reported in red while WAVE.th is reported in green.

Case A: example of healthy lung from dataset 1. The Gaussian fit is well overlapped to the CT relative histogram, with a slight deviation on the right side of the black curve, caused likely by the inclusion in the lung segmentation of major vessels. Case B: example of patient with cancer in the left lung, from dataset 2. The presence of the tumor mass and the consequent distortion of the surrounding structures increase the deviation between the Gaussian fit and the CT relative histogram. Case C: COVID-19 patient with an extended ground glass opacification of the lung that enlarges the Gaussian fit. In addition to the fact that Mu.f value is on the right side of the green window (-950/-700 HU), it causes an important difference between the two WAVE metrics. The lungogram shows that the higher values of the biomarker $\Delta HU_{Mu\text{-}Avg}$ are found in the dorsal regions. Case D: Widespread disease with only some portions of healthy lung, but sufficient to produce a suitable histogram for Gaussian fit. In this case, WAVE.f, and WAVE.th calculations returned similar results even if Mu.f value is not in the central position of the range used to calculate WAVE.th. The lungogram shows that the higher values of $\Delta HU_{Mu\text{-}Avg}$ are found in the dorsal regions.

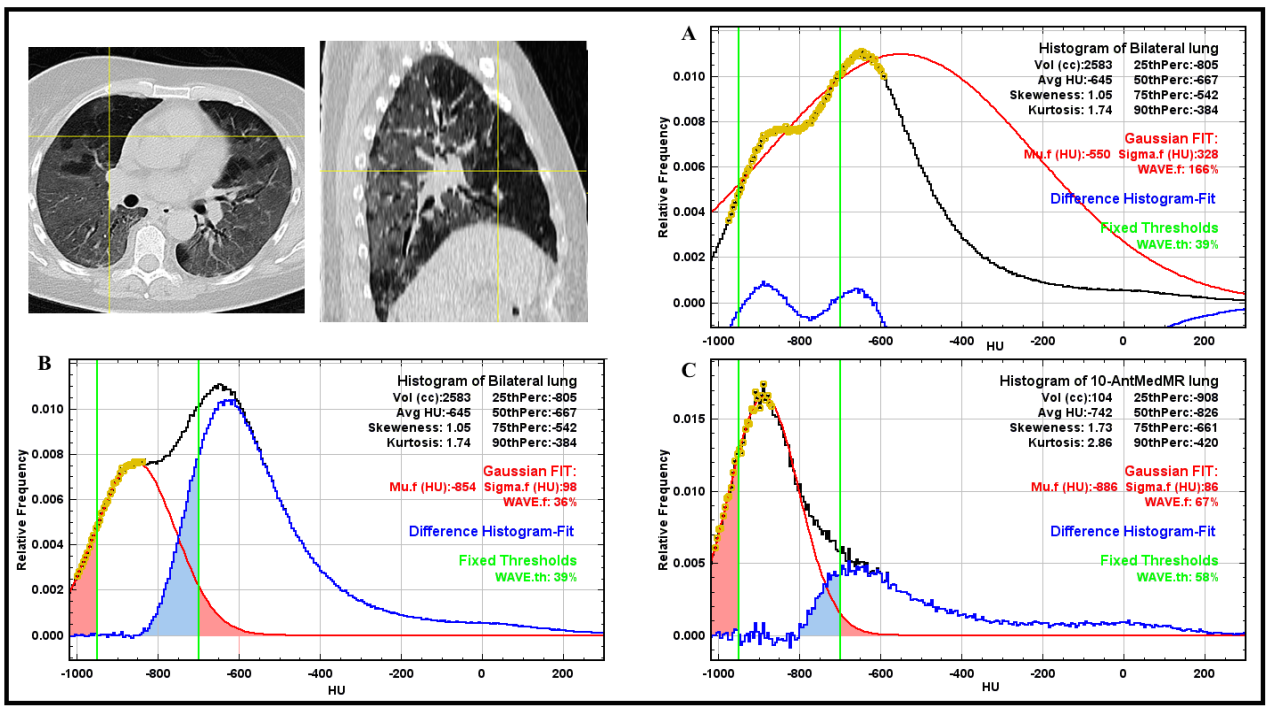

Fig. 6: Example a COVID patient with widespread opacification, displayed in axial and sagittal views. Plot A: the modal value in the CT lung histogram is greater than -700 HU and the definition of the fit range based on this value is not correct for the Gaussian model. Plot B: Data points for Gaussian fit are defined by the local maximum identified by first derivative. Plot C: Analysis performed in the anterior-medial of the central portion of the right lung where, as showed by the yellow cross-hair in the CT images, normal lung is prevalent. The light red shadowed areas under the curves represent the lung volume classified as "well aerated" according to the Gaussian model but not according to theWAVE.th definition. Analogously, the light blue shadowed areas under the curves represent the lung volume classified as "well aerated" according to the fixed threshold model but not according to the Gaussian metric. The difference between the two metrics arise when these two areas do not offset each other.

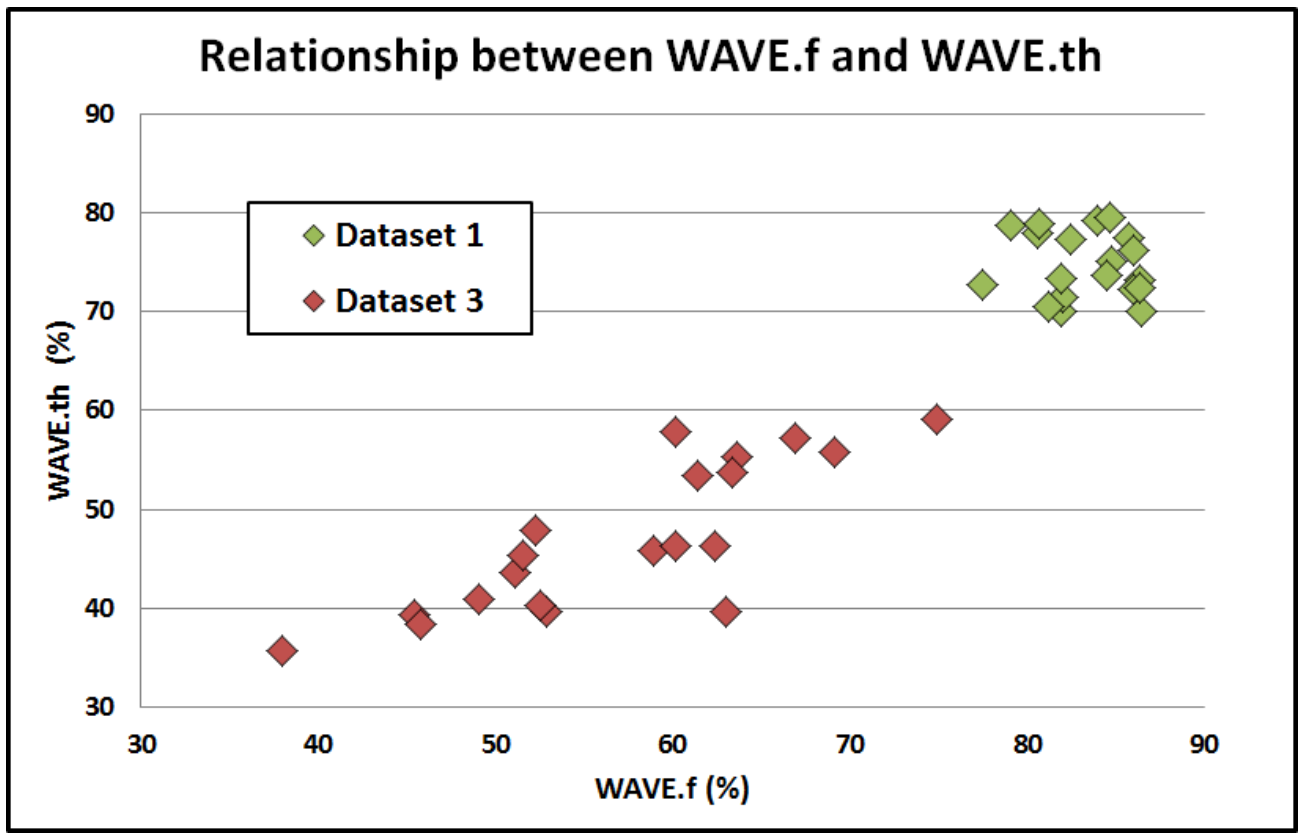

Fig. 7:Scatterplot of WAVE.f and WAVE.th values for patients in dataset 1 (healthy lung) and dataset 3 (COVID-19 patients).